**Gate-tunable Exchange Bias and Voltage-controlled Magnetization Switching in a van der Waals Ferromagnet**

*Mayank Sharma, Garen Avedissian, Witold Skowroński, Junhyeon Jo, Andrey Chuvilin, Fèlix Casanova, Marco Gobbi and Luis E. Hueso\**

M. Sharma, Dr. G. Avedissian, Dr. J. Jo, Prof. F. Casanova, Prof. A. Chuvilin, Dr. M. Gobbi, Prof. L. E. Hueso
CIC nanoGUNE, 20018 Donostia-San Sebastian, Basque Country, Spain
E-mail: m.sharma@nanogune.eu l.hueso@nanogune.eu

Prof. A. Chuvilin, Prof. F. Casanova, Dr. M. Gobbi, Prof. L. E. Hueso
IKERBASQUE, Basque Foundation for Science, 48009 Bilbao, Basque Country, Spain

Prof. W. Skowroński
AGH University of Krakow, Institute of Electronics, Al. Mickiewicza 30, Kraków 30-059, Poland

Dr. M. Gobbi
Centro de Física de Materiales (CFM-MPC) Centro Mixto CSIC-UPV/EHU, 20018 Donostia-San Sebastián, Basque Country, Spain

Abstract:

The discovery of van der Waals magnets has established a new domain in the field of magnetism, opening novel pathways for the electrical control of magnetic properties. In this context, $Fe_3GeTe_2$ (FGT) emerges as an exemplary candidate owing to its intrinsic metallic properties, which facilitate the interplay of both charge and spin degrees of freedom. Here, we demonstrate the bidirectional voltage control of exchange bias (EB) effect in a perpendicularly magnetized an all-van der Waals FGT/O-FGT/hBN heterostructure. The antiferromagnetic O-FGT layer was formed by naturally oxidizing the FGT surface. The observed EB magnitude reaches 1.4 kOe with a blocking temperature (150 K) reaching close to the Curie temperature of FGT. Both the exchange field and the blocking temperature values are among the highest in the context of layered materials. The EB modulation exhibits a linear dependence on the gate voltage and its polarity, observable in both positive and negative field cooling (FC) experiments. Additionally, we demonstrate gate voltage-controlled magnetization switching, highlighting the potential of FGT-based heterostructures in advanced spintronic devices. Our findings display a methodology to modulate the magnetism of van der Waals magnets offering new avenues for the development of high-performance magnetic devices.

## 1. Introduction

Exchange bias (EB) is a phenomenon observed when an antiferromagnetic (AFM) material is in contact with a ferromagnetic (FM) or ferrimagnetic material. This proximity creates a unidirectional magnetic anisotropy at the interface, observed as a horizontal shift in the magnetic hysteresis loop of the FM when subjected to magnetic field cooling.[1] EB was first observed in Co/CoO core-shell nanoparticles and, since then, it has been extensively studied in thin films owning to its wide range of commercial applications in permanent magnets, magnetic recording media, and magnetic random-access memories (MRAM).[2–4]

Although EB has been vital to spintronic applications, further integration into more advanced devices may necessitate external control of the effect. In this respect, voltage control has emerged as the leading candidate as it provides unparalleled advantages. These include lower heat dissipation when compared to current controlled devices as well as reduced interference with neighboring magnetic elements due to absence of stray fields, which are commonly generated by magnetic control methods.[5] Consequently, voltage control allows for device miniaturization and integration with existing technology.[6–10] This has attracted researchers towards its study in different thin film systems, for instance in CoPd/$Cr_2O_3$, in Co/Pt/IrMn and, more recently, in self-oxidized Co/CoO.[11–13]

The discovery of two-dimensional (2D) van der Waals (vdW) materials and, particularly, 2D magnets such as $CrI_3$, $Cr_2Ge_2Te_6$, $Fe_3GeTe_2$ (FGT), $CrTe_2$, CrSBr, or $Fe_3GaTe_2$ have opened vast new opportunities for spintronic devices.[14–20] In addition, heterostructures based on 2D materials offer unprecedented stacking possibilities, no interlayer mixing, and emerging physical properties with possibility of modulation with external stimuli e.g., gate voltage.

FGT and FGT-based heterostructures have gained special attention due to its strong out-of-plane magnetic anisotropy, relatively high Curie temperature, and metallic nature, allowing for the interplay of both charge and spin degree of freedom.[21,22] These properties led to exploration of EB in FGT via different approaches such as molecular spinterface effect, strain modulation, natural oxidation and the combination of FGT and 2D AFM.[23–26]

Despite significant advancements, the control of EB in van der Waals heterostructures remains largely underexplored.[27–29] Furthermore, the potential to utilize an electric field effect to reverse the magnetization in van der Waals magnets remains elusive. In this work, we report bidirectional electrical gate control of EB and deterministic magnetization switching in an all-van der Waals FGT/O-FGT/hBN heterostructure. The EB has a large magnitude of 1.4 kOe with a blocking temperature (150 K) close to the Curie temperature of FGT. Both the exchange field and the blocking temperature values are among the highest in the context of layered materials (see supporting table S1).[24,26,30–37] The bidirectional control of EB is achieved with a 2D solid-state dielectric material composed of insulating hexagonal boron nitride (hBN). The EB modulation is present on both positive and negative field cooling (FC) experiments and shows a linear dependence on the gate voltage ($V_g$) and its polarity. Furthermore, we engineer the EB modulation to demonstrate a deterministic voltage-controlled magnetization switching in our heterostructures, highlighting the potential of FGT-based heterostructures in advanced spintronic devices.[38,39] Our findings present a methodology to modulate the magnetism of van der Waals magnets offering new avenues for the development of high-performance voltage controlled magnetic devices.

**2. Results and discussion**

We mechanically exfoliate FGT flakes with a viscoelastic stamping made of polydimethylsiloxane (PDMS) in an argon-filled glovebox and transfer it onto Si/SiO$_2$ substrates with patterned Au metallic electrodes. The so-obtained sample was transferred to a hot plate outside the glovebox and annealed at 100 ºC for 30 minutes. This method was previously demonstrated to generate an oxide layer on FGT.[24] Finally, the sample is encapsulated with hBN (see Methods for details). This material has a two-fold purpose: to stop further oxidation and to act as the top-gate dielectric. We use hBN due to its excellent performance owing to its nearly perfect crystallinity with minimum pin holes.[40,41] This results in extremely low level of leakage currents when compared to sputtered oxide gate on a 2D magnet.[27] The final device stack is illustrated in **Figure 1**a.

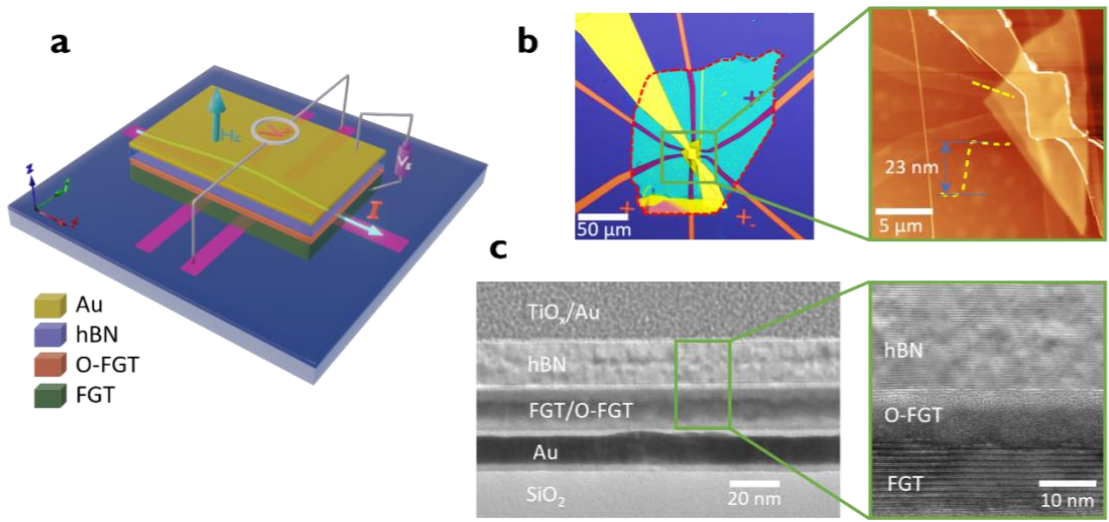

**Figure 1.** Device schematic and different microscope images. a) Device sketch, an electrical current ($I$) is applied along the longitudinal direction ($x$) and the Hall voltage ($V_{xy}$) is measured transversally (along $y$) while the magnetic field is applied along the out-of-plane direction ($z$). b) Top view optical image of the device. The inset shows an atomic force microscope image of the device. The FGT flake, bottom Au contacts, and the top gate are visible. c) Cross-sectional TEM images of the device with a zoom of a high-resolution image of the 2D heterostructure.

**Figure 1**b shows an optical image of the device, while an atomic force microscope image is in the inset. To gain an in-depth insight on the formed oxide layer and of the quality of the interface, we utilized transmission electron microscopy (TEM). **Figure 1**c shows the cross-sectional image of the full heterostructure on the Si/SiO$_2$ substrate. The inset is the high-resolution image of the device focused on the layered FGT structure along with the naturally formed oxide (O-FGT) and the top hBN. A clean interface is clearly observed between the FGT and its oxide. The electron energy loss spectroscopy (EELS) data for the heterostructure is presented in the supporting Figure S1.

To investigate the effect of the O-FGT on the magnetic properties of the FGT, we study its anomalous Hall effect (AHE). This provides us with information on the magnetic interactions at the FGT/O-FGT interface. First, we perform the magnetic field cooling (FC) from 300 K to 10 K with the application of ±6 kOe out-of-plane magnetic field ($H_z$). Once cooled down, we apply an electric current along the x axis of the FGT device and we sweep the magnetic field in an out-of-plane orientation to record the Hall voltage (along y).

**Figure 2**a shows the hysteresis loops at a representative temperature of 20 K. The loop shifts towards a positive (negative) field magnetic direction when measured with a negative (positive) FC, resulting in an exchange bias field ($H_{EB}$) of 1.1 kOe. **Figure 2**b shows the AHE loops at four different temperatures when measured with positive FC. The $H_{EB}$ is negative and its absolute value decreases with increasing temperature. The AHE loops maintain the square shape resulting from coherent single domain switching at all the temperatures, even after the process of oxidation. In **Figure 2**c, we plot the extracted $H_{EB}$ from both the positive and negative FC measurements. The highest $H_{EB}$ (~1.4 kOe) is observed at 10 K, while it disappears at a blocking temperature of 150 K. The EB observed in our samples is attributed to the antiferromagnetic character of the O-FGT layer in the heterostructure.[24]

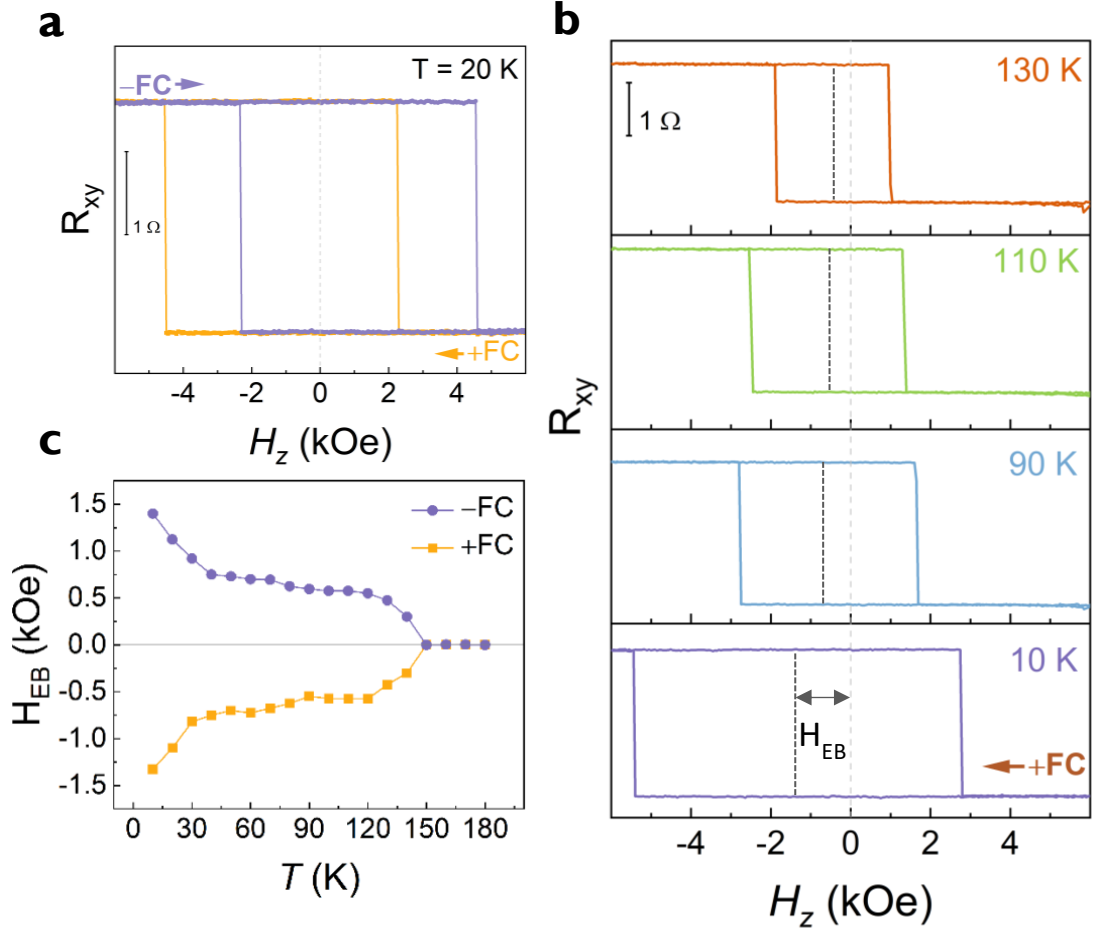

**Figure 2.** Exchange bias in the FGT/O-FGT heterostructure. a) Hall resistance loop at a representative temperature of 20 K measured after positive FC (+6 kOe, orange line) and negative FC (–6 kOe, purple line). The magnetic hysteresis associated with the AHE of FGT is shifted by identical magnitude in opposite directions with respect to the polarity of the FC. b) Hall resistance loops measured after a positive FC (+6 kOe) at selected temperatures. The hysteresis loops show a decrease in the asymmetry as the temperature increases. c) Exchange bias field ($H_{EB}$) extracted from the Hall resistance loops as a function of temperature, $H_{EB} = \frac{H_{C+} - H_{C-}}{2}$ where $H_{C+}$ is the positive coercive field and $H_{C-}$ is the negative coercive field. The positive exchange bias (measured with negative FC) is depicted in purple and negative exchange bias (measured with positive FC) in orange.

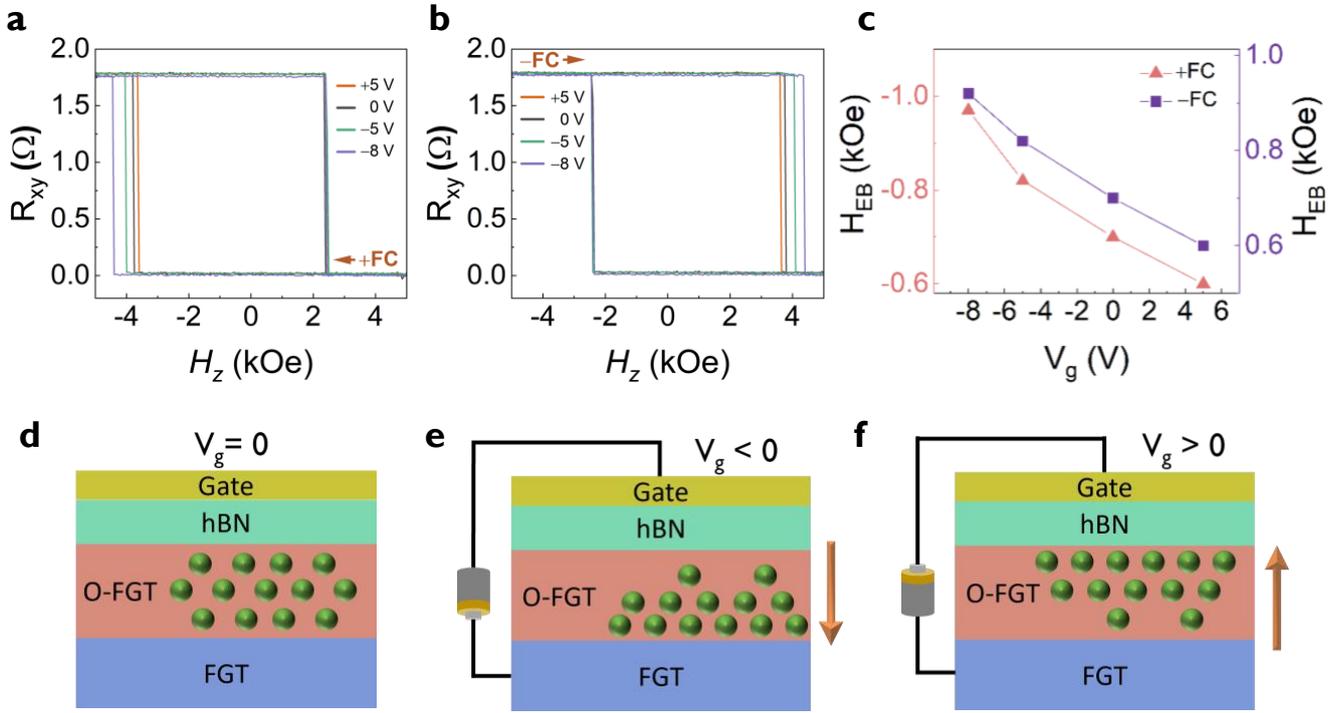

**Figure 3.** Gate voltage control of EB. a) Hall resistance loops measured at 60 K after positive FC (+6 kOe) and at different gate voltages ($V_g$): +5, 0, −5 and −8 V. b) Hall resistance loops measured at 60 K after negative FC (−6 kOe) and at different $V_g$: +5, 0, -5 and -8 V. c) Gate dependence of HEB with positive and negative FC. d) Schematic representation of the distribution of oxygen ions within the oxide layer under $V_g = 0$ V, where the ions are uniformly distributed throughout the entire layer. e) Under a negative gate voltage ($V_g < 0$ V), oxygen ions accumulate closer to the FGT/O-FGT interface, leading to stronger exchange coupling. f) Under a positive gate voltage ($V_g > 0$ V), oxygen ions move away from the interface, weakening the exchange coupling. The orange arrows indicate the direction of motion of the oxygen ions in response to the applied gate voltage.

We then perform magnetotransport measurements on the FGT/O-FGT heterostructure with different electrical gate voltages using the top gate configuration shown in Figure 1a.

**Figure 3** depicts the electrical gate voltage control of exchange bias in our heterostructure. Both for positive and negative (**Figure 3**a and b) FC, we observe a clear shift in the hysteresis loop upon the application of different gate voltages ($V_g$). The modulation of the exchange bias is mediated by the movement of oxygen ions, which either accumulate near to the interface under $V_g < 0$ V (**Figure 3**e) or move away under $V_g > 0$ V (**Figure 3**f), thereby altering the density and distribution of pinning sites. The effect will be further elaborated upon in the subsequent section. In both field cooling experiments, only one branch of coercivity, i.e. positive $H_{C+}$ (negative $H_{C-}$) in case of the negative (positive) FC, changes while the other branch remains relatively constant. Similar behavior has been observed in other FM/AFM systems, where variations in the FM thickness or temperature induce analogous effects.[42–44] Such an effect is also possible with gating as seen in our system, mediated by the change in the ratios of anchored and unanchored uncompensated. The extracted $V_g$ dependence of the EB is depicted in **Figure 3**c, which shows a linear dependence on the applied $V_g$.

Our results are reproducible across different samples, as shown in supporting Figure S2. To further confirm the critical role of the O-FGT layer in modulation of the exchange bias, we prepared a control device without the O-FGT layer. This control device exhibited no exchange and no detectable change in the coercivity of FGT under similar gate voltages (supporting information Figure S3). These results confirm that the observed modulation of exchange bias is directly attributable to the presence of the O-FGT layer. Furthermore, during the electrical gating experiments, we recorded a leakage current of less than 70 pA (supporting information Figure S4), which is five orders of magnitude lower than the injected measurement current, eliminating Joule heating as a possible source of the observed effect.

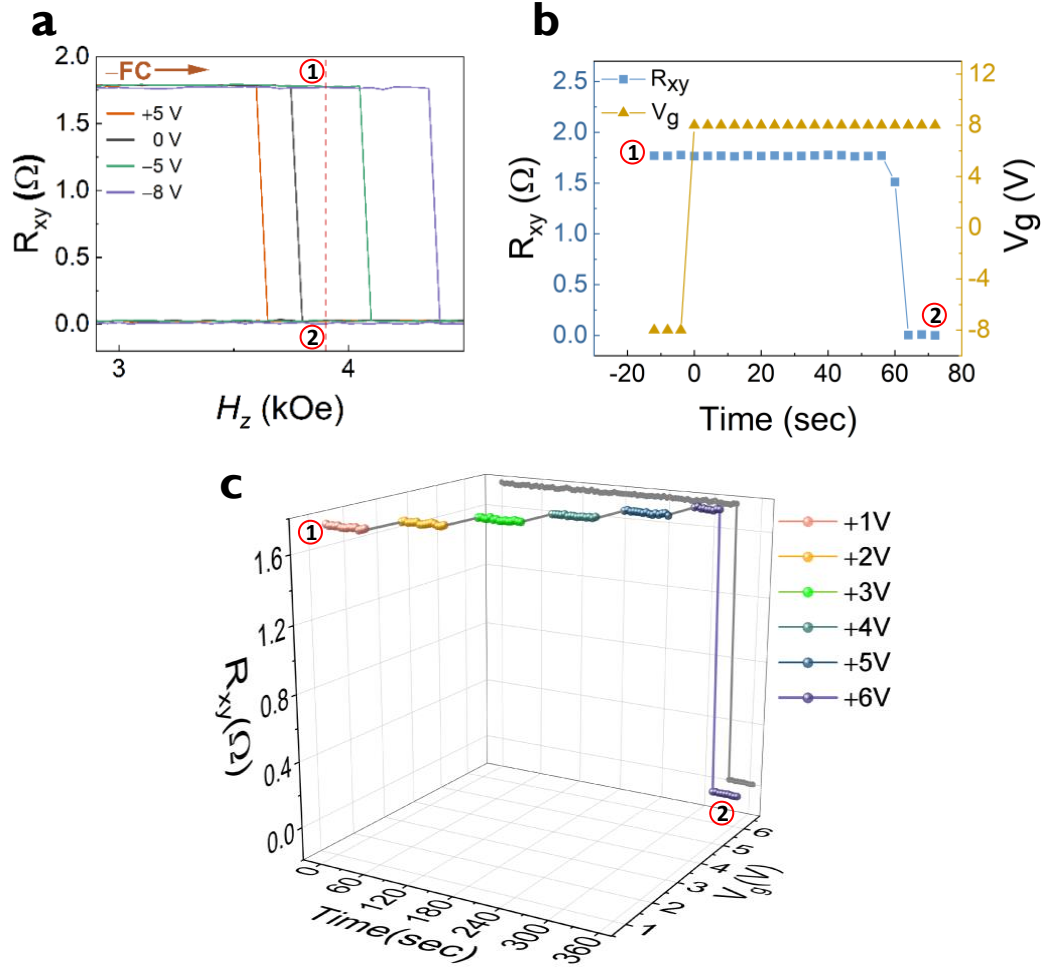

**Figure 4.** Magnetization switching with gate. a) Cropped view of the AHE measured at 60 K after negative FC (−6 kOe) and at different gate voltages ($V_g$): +5, 0, −5, −8 V. The red dashed line at 3.9 kOe represents a field between the EB switching fields corresponding to $V_g$ = 0 V and $V_g$ = −5 V. The numbers 1 and 2 (in red circles) represent the upward and downward magnetization state, respectively. b) Magnetization switching induced by reversing gate voltage polarity. The measurement is performed after initialization of system in high EB state with application of $V_g$ = −8 V and bringing it to state 1. The gate voltage polarity is switched to $V_g$ = +8 V at t = 0 resulting in magnetization switching of the FGT, observed by a drop in the Hall resistance from 1 to 2 at t = 60 s. c) Magnetization switching with gate voltage sweep. Similar initialization of state 1 with $V_g$ = −8 V followed by a gate voltage sweep performed with application of each voltage for 60 seconds in steps of 1 V starting from +1 V. The magnetization is switched with the application of $V_g$ = +6 V again seen by Hall resistance drop to state 2.

Utilizing the control of EB, we demonstrate magnetization switching via the application of an electrical gate voltage. For this experiment, we select a magnetic field (denoted by red dashed line in **Figure 4**a) that allows access to both upward and downward magnetization states, denoted by points 1 and 2, respectively (**Figure 4**a). The system is initially set to upward magnetization and high EB state by performing negative FC and applying a negative gate voltage. Upon reversing the gate voltage polarity, the system transitions from the high EB state to the low EB state. This transition is driven by the movement of oxygen ions away from the interface under the influence of the positive gate voltage (**Figure 3**f). Since the magnetic field is parked between these two states, this reduction in exchange bias also induces a change in the magnetization of the system, causing it to switch from the upward to downward state. As illustrated in **Figure 4**b, the Hall resistance of FGT, and consequently the magnetization of the system, switches deterministically from the upward to the downward state when the gate voltage polarity is switched from -8 V to +8 V. Furthermore, an upward magnetization switch from downward state is also possible by utilizing a positive FC process, as demonstrated in supporting information Figure S5. This shows that both upward and downward magnetization states can be controlled by varying the gate voltage and FC conditions.

As a control experiment, we apply $V_g = -8$ V for a similar duration of time, and we observe no change in the Hall resistance of the FGT (see supporting Figure S6a). Finally, we perform the magnetization switching experiment with gate voltage sweep. In **Figure 4**c, the voltage is swept in steps of 1 V, and we again see that the magnetization switches from upward to downward state at $V_g$ = +6 V. In this case, for the control experiment, we perform a similar gate voltage sweep with smaller time steps and we do not observe a switch until the voltage reaches +6 V (see supporting Figure S6b). These two control experiments eliminate the possibility of magnetization switching arising from thermal fluctuations. A further control experiment can be found in supporting Figure S7, where a sputtered insulator is used as gate dielectric, leading to a high leakage current and to magnetization switching due to thermal fluctuations.

The experiments on magnetization switching via gate voltage, conducted under both constant voltage and voltage sweep conditions, demonstrate that the magnetization state does not undergo immediate state transition upon the reversal of gate voltage polarity. This response is a significant departure from the rapid changes one would expect if electrostatic charge doping were the driving mechanism.[45–47] In typical electrostatic gating scenarios, the time scale of charge accumulation or depletion at the interface can range from picoseconds to nanoseconds. This extremely fast dynamics results from the high mobility and swift response of electronic charges. Hence our finding effectively rules out electrostatic charge doping as a potential dominant mechanism for the observed EB modulation.

Our observations of low leakage current also effectively eliminate Joule heating as a cause of the observed EB modulation. Additionally, the response time in gate voltage-controlled magnetization switching experiments rules out a purely electrostatic gating mechanism. These findings compel us to consider a more gradual diffusion-based mechanism, specifically, the migration of oxygen ions in the O-FGT layer.[48] Oxygen ion migration, characterized by its relatively slow dynamics in comparison to electronic processes (usually spanning from tens to hundreds of seconds under the influence of electric field strength of ~$MVcm^{-1}$) offers a plausible explanation for the gradual modulation of magnetic properties in our system.[49–51]

Oxygen ion migration influences the exchange bias by altering the interface characteristics with different gate voltages, particularly affecting the density and distribution of pinning sites, thereby modifying the exchange interaction. The enhancement in exchange bias with negative gate voltages could be attributed to the movement of oxygen ions closer to the FGT interface, resulting in an increase in anchored uncompensated spins within the AFM. However, this migration could also increase the number of unanchored uncompensated spins. Thus, oxygen migration would impact two mechanisms: the spins responsible for the EB and the spins that increase frustration at the AFM-FM interface, contributing to increased coercivity.[52] These two mechanisms together would generate an increase in $H_{EB}$ with negative gate voltage while keeping one branch of the coercivity relatively constant. Conversely, the application of positive gate voltages creates oxygen vacancies at the interface with less pinning sites, and hence would result in a lower $H_{EB}$.

The observed voltage control of EB can be quantitatively explained by the following model. We can estimate the voltage drop within the hBN layer ($V_{hBN}$) and across the O-FGT layer ($V_{OFGT}$) by utilizing boundary condition at the interface. In a continuous model, the displacement normal to the interface, i.e. D = εE, remains uniform across both sides of the interface. By utilizing the boundary condition at the interface,

$$\frac{\varepsilon_{OFGT} V_{OFGT}}{t_{OFGT}} = \frac{\varepsilon_{hBN} V_{hBN}}{t_{hBN}} \tag{1}$$

where $t_{OFGT}$ is the thickness of the O-FGT layer and $t_{hBN}$ is the thickness of the hBN. Since $V_{OFGT} + V_{hBN} = V_g$ and $\varepsilon_r = \frac{\varepsilon_{hBN}}{\varepsilon_{OFGT}}$, we can find the voltage drop across the O-FGT layer as

$$V_{OFGT} = \frac{t_{OFGT} \varepsilon_r V_g}{t_{OFGT} \varepsilon_r + t_{hBN}} \tag{2}$$

Considering $V_g$ to be 8 V, $t_{OFGT}$ as 5 nm, $t_{hBN}$ as 18 nm, and $\varepsilon_r$ as 0.19, the calculation yields a $V_{OFGT}$ of 0.4 V which translates into an electric field of 4 MVcm$^{-1}$.[40,53] This considerable electric field strength within the oxide layer effectively overcomes the electronic barrier and enables the migration of oxygen ions (typically <2 eV).[54,55] Therefore, we can now proceed to evaluate the mobility of oxygen ions.

The total mobility (μ) can be expressed as a sum of a temperature-independent part $μ_0$ and a temperature-dependent part. The temperature-dependent part often follows an Arrhenius relation:

$$μ(T) = μ_0 + μ_{th} exp\left(\frac{-E_a}{k_B T}\right) \qquad (3)$$

where $μ_0$ and $μ_{th}$ are coefficients for temperature independent and dependent parts of mobility, respectively. $E_a$ is the activation energy for ion migration, $k_B$ is the Boltzmann constant, and $T$ is the temperature. The average value for $μ_0$ is $10^{-15}$ cm$^2$V$^{-1}$s$^{-1}$ and for the $μ_{th}$ is $10^{-13}$ cm$^2$V$^{-1}$s$^{-1}$.[56] Since the experiment is performed at 60 K, the thermal contributions to the mobility are negligible and, therefore, the mobility expression reduces to $μ_0$. Hence, the distance covered by ions is $μ_0 E t$, where $μ_0 E$ is the ion drift velocity ($v$) and $t$ is the time under gate voltage application. Using the above expression, we get $v$ = 0.4 Ås$^{-1}$ and distance cover as 2.4 nm in 60 seconds. This interpretation provides a plausible account of the observed voltage control of EB while allowing us to present a well-rounded quantitative understanding of the dynamic behavior observed in magnetization switching.

## 3. Conclusions

The present work offers a comprehensive and detailed examination of the EB phenomenon within the 2D van der Waals ferromagnet FGT, featuring a FGT oxide layer (O-FGT). This oxide layer, as evidenced by high-resolution TEM, maintains a pristine interface with the underlying FGT material and exhibits AFM properties. These properties are instrumental in pinning the ferromagnetic FGT, resulting in a pronounced $H_{EB}$ of 1.4 kOe. It is observed that this $H_{EB}$ diminishes with an increase in temperature and vanishes at a blocking temperature of 150 K. Additionally, this study elucidates the modulation of EB through electrical gating, employing the 2D insulator hBN as a gate dielectric. Our findings reveal a pronounced dependence of the EB effect on both the magnitude and polarity of the applied $V_g$. Utilizing voltage control, we have achieved deterministic magnetization switching, marking a significant stride in the field of spintronics by integrating the properties of van der Waals materials with advanced functional control over magnetic phenomena

4. Experimental section

Sample preparation:

Ti (2 nm)/Au (11 nm) prepatterned electrodes with Hall-bar (channel dimension 10 µm x 2 µm) geometry were prepared on Si/SiO$_2$ substrate. A FGT flake (23 nm) (HQ graphene) was prepared by mechanical exfoliation using a blue tape (Nitto® SPV224) and transferred to a PDMS film (Gel-pak).

The exfoliated FGT flake thickness on PDMS film was characterized using the optical contrast in an in situ optical microscope and was further confirmed with the atomic force microscopy (Agilent 5500 SPM). The desired FGT flake was transferred onto the electrode-prepatterned substrate using standard dry transfer stamping technique. The exfoliation and stamping were performed inside an argon atmosphere glovebox with H$_2$O and O$_2$ < 0.1 ppm.

To perform the natural oxidation the FGT sample was annealed in atmospheric conditions on a hot plate for 30 minutes at 100 °C. The sample was transferred back to the glovebox where the encapsulation with hBN (18 nm) (HQ graphene) was done as described before. The top gate on the sample was prepared using positive lithography followed by Ti/Au electrode deposition and lift-off process.

TEM analysis:

TEM and STEM data was acquired using TitanG2 60-300 electron microscope (FEI, Netherlands) equipped with xFEG, monochromator, image aberration corrector, HAADF STEM detector and Quantum GIF (Gatan, UK). Images were obtained at 300kV accelerating voltage. Cross-section samples of the devices have been prepared by a standard FIB protocol using Helios 600 FIB/SEM (FEI, Netherlands).

Magneto transport measurement:

Electrical transport measurements were performed in a physical property measurement system (PPMS, Quantum Design) with a rotational sample stage. An electrical current was applied by a Keithley 6221, and a measured voltage was detected by a Keithley 2182 nanovoltmeter. The transverse voltage was recorded with DC reversal (delta) method. This involves recording the voltage with the current reversal leading to elimination of thermal signal contributions from the measurement. For the field cooling (FC) procedure, an out-of-plane magnetic field of ±6 kOe was applied while the sample was cooled from 300 K to the target measurement temperature. For applying the gate voltage, a Keithley 2636 source meter was utilized.

**Acknowledgements**

We thank Dr. Rafael Morales for helpful discussions. We acknowledge funding by the Spanish MCIN/AEI/10.13039/501100011033 and by ERDF A way of making Europe (projects No. PID2021-122511OB-I00, PID2021-128004NB-C21 and Maria de Maeztu Units of Excellence Programme No. CEX2020-001038-M) and by the European Union H2020 Programme (projects No. 955671-SPEAR, No. 965046-INTERFAST and No. 964396-SINFONIA). This work was also supported by the FLAG-ERA grant MULTISPIN, via the Spanish MCIN/AEI with grant number PCI2021-122038-2A. J.J. acknowledges the Spanish MCIN/AEI and the European Union NextGenerationEU/PRTR for a "Juan de la Cierva" fellowship (grant No. FJ2020-042842-I). W.S. acknowledges grant No. BPN/BKK/2022/1/00010/U/00001 from Polish National Agency for Academic Exchange. M.G. acknowledges support from the "Ramón y Cajal" Program by the Spanish MCIN/AEI (grant No. RYC2021-031705-I).

Supporting Information

**Gate-tunable Exchange Bias and Voltage-controlled Magnetization Switching in a van der Waals Ferromagnet**

*Mayank Sharma, Garen Avedissian, Witold Skowroński, Junhyeon Jo, Andrey Chuvilin, Fèlix Casanova, Marco Gobbi and Luis E. Hueso\**

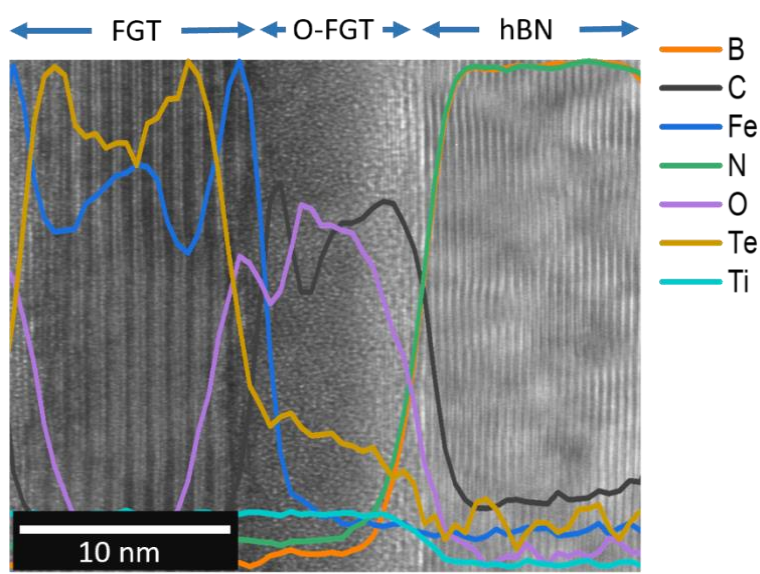

**Figure S1.** Electron energy loss spectroscopy (EELS) plot showing normalized intensities. An oxygen-rich layer is clearly visible between the FGT and the hBN, corresponding to the O-FGT layer. The boron and nitrogen peaks at the end indicate the top hBN layer. Additionally, a carbon peak is observed between the FGT and hBN, attributed to PDMS polymer residues on the FGT flake from the stamping process.

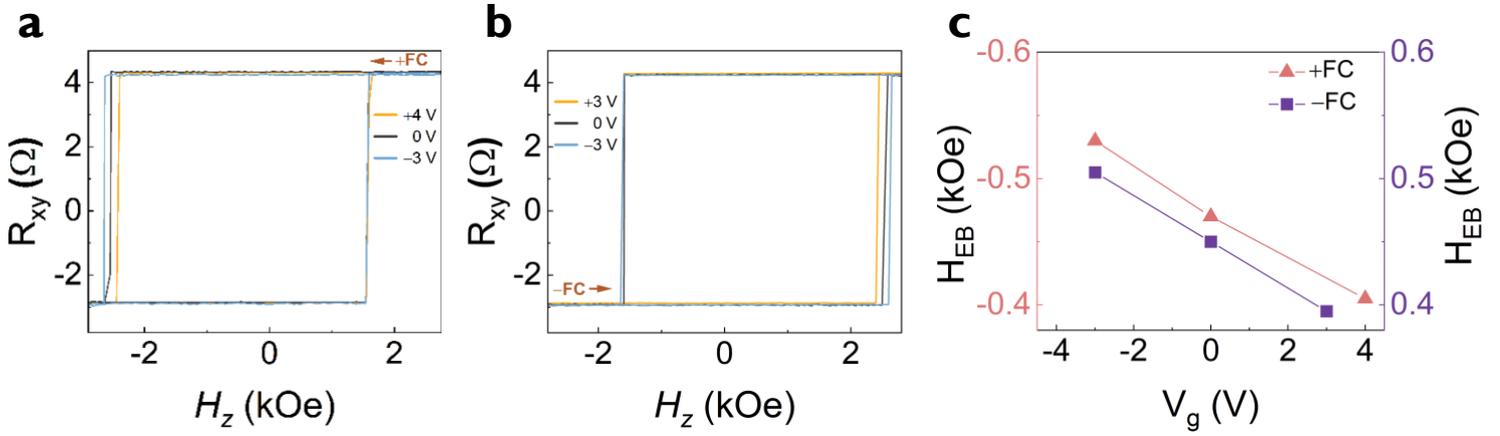

**Figure S2.** Gate voltage control of EB in sample B a) Hall resistance measured after positive FC (+4 kOe) to 40 K as a function of different gate voltages +4, 0 and -3 V. The EB magnitude increases with the negative gate voltage and decreases with the positive gate voltage. b) Hall resistance measured after negative FC (-4 kOe) at different gate voltages +3, 0 and -3 V. c) Voltage dependence of EB with positive and negative FC. Linear dependence on applied gate voltage is observed with both positive and negative FC measurements.

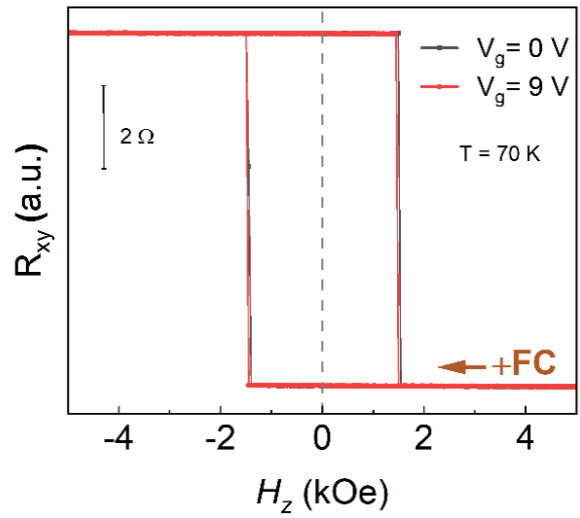

**Figure S3.** Control device experiment without the O-FGT layer. The device had the same geometry, including hBN encapsulation and a top gate, as used in the main study. Under field cooling of 6 kOe no exchange bias is observed. No detectable change in the coercivity of the FGT is observed under gate voltage, the AHE loops in black ($V_g$ = 0 V) and red ($V_g$ = 9 V) overlap.

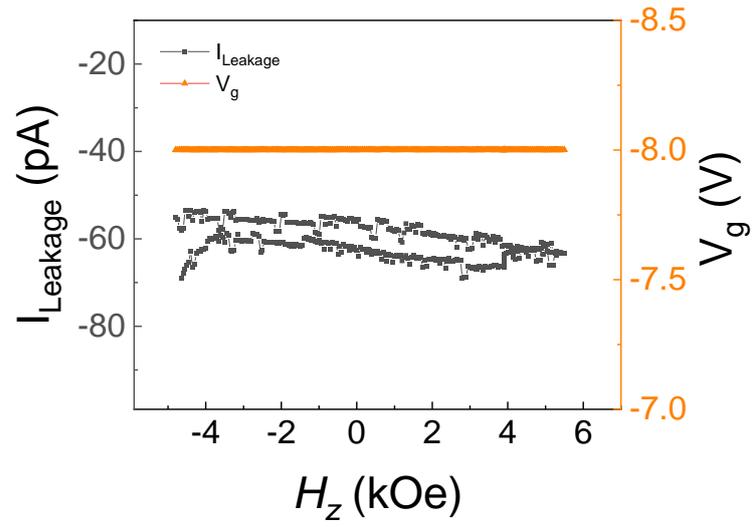

**Figure S4.** Leakage current through the gate with the application of -8 V during the whole field sweep performed after the negative filed cooling, the leakage current recorded is below 70 pA.

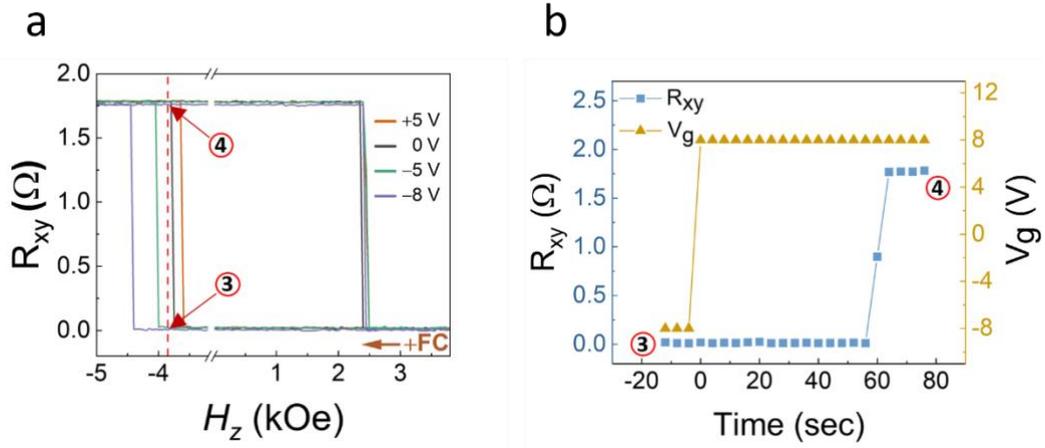

**Figure S5.** Magnetization switching from the downward to the upward state. a) Cropped view of the AHE measured at 60 K after positive FC (+6 kOe) and at different gate voltages ($V_g$): +5, 0, −5, −8 V. The red dashed line at -3.8 kOe represents a field between the EB switching fields corresponding to $V_g = 0$ V and $V_g = -5$ V. The numbers 3 and 4 (in red circles) represent the downward and upward magnetization state, respectively. b) Magnetization switching induced by reversing gate voltage polarity. The measurement is performed after initialization of system in downward magnetisation and high EB state with application of $V_g = -8$ V and bringing it to state 3. The gate voltage polarity is switched to $V_g = +8$ V at t = 0 resulting in magnetization switching of the FGT, observed by an increase in the Hall resistance from 3 to 4 at t = 60 s

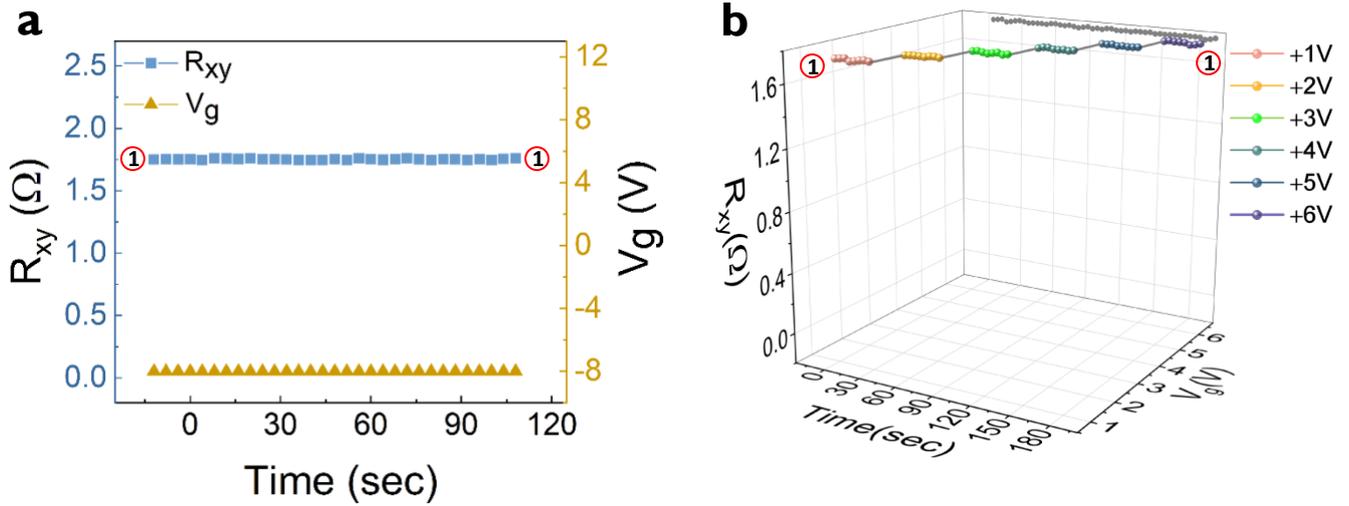

**Figure S6.** Control experiment for the magnetization switching with the gate voltage. a) Measured after initialization of system upward EB state with application of -8 V and bringing it to state 1. The gate voltage is maintained at -8 V and the system is maintained at state 1. No magnetization switching is observed. c) Similar initialization of state 1 with -8 V gate followed by a gate volage sweep performed with application of each voltage for 30 seconds in steps of 1 V starting from +1 V. No magnetization switching is observed the system maintains state 1.

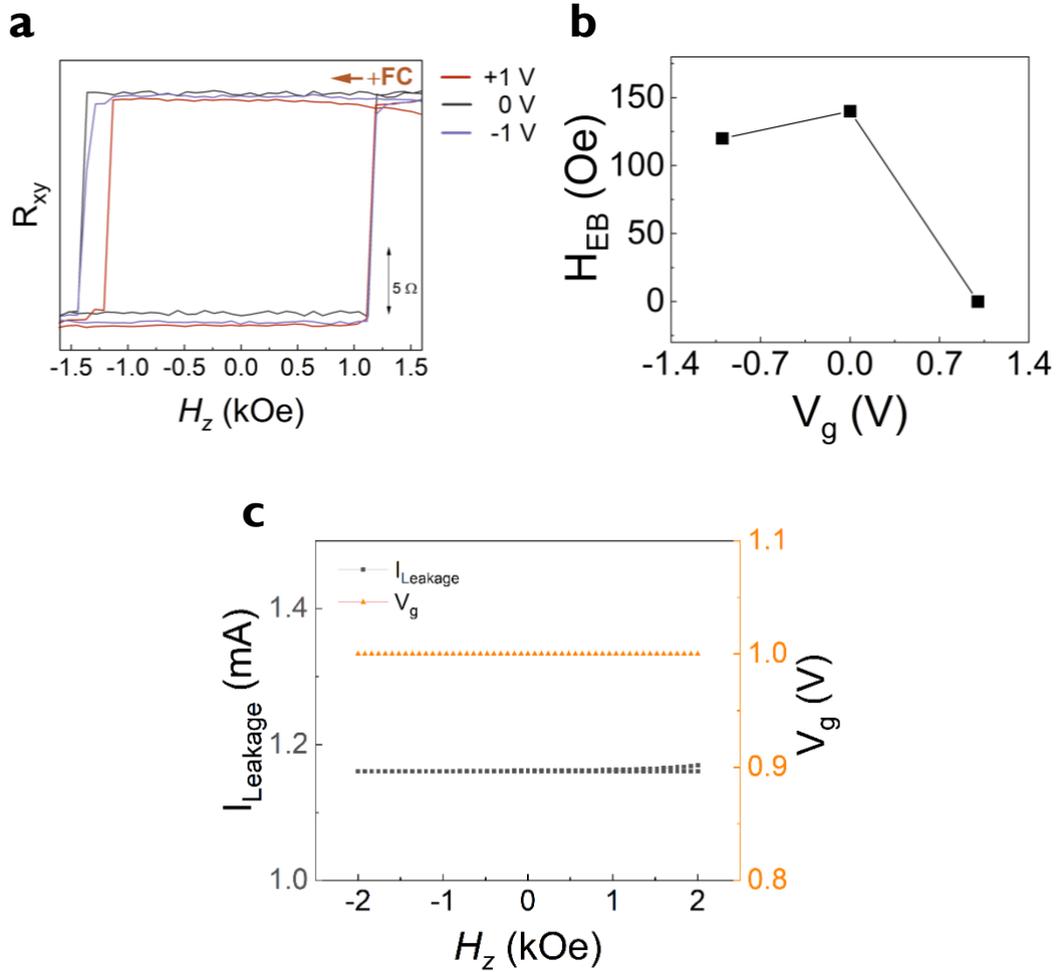

**Figure S7.** Gate voltage control of EB in sample C a) Hall resistance measured after positive FC (+1.5 kOe) to 40 K as a function of different gate voltages +1, 0 and -1 V. b) EB as a function of gate voltage. EB magnitude has little modulation with the application of negative gate voltage and ceases to exist with the application of positive 1 V. c) Leakage current with the application of 1 V of gate voltage, measured during a magnetic field sweep. The leakage current is higher than the applied measurement current.

| Structure | Highest EB (Oe) | Reference |
|---|---|---|
| FGT/MnPSe$_3$ | 233 | [30] |
| FGT/CrCl$_3$ | 560 | [31] |
| FGT/FePS$_3$ | 260 | [32] |
| FGT/CrOCl | 400 | [33] |
| FGT/MnPS$_3$ | 255 | [34] |
| FGT/IrMn | 895 | [35] |
| FGT/O-FGT | 1400 | This work |
| FGT/i-FGT | 474 | [27] |
| FGT/CrSBr | 1500 | [37] |
| FePSe$_3$/FGT | 1112 | [36] |
| FGT/o-FGT | 493 | [24] |
| FGT/CoPc | 840 | [26] |

**Table S1**. Highest EB reported in different FGT based systems.